\documentclass[runningheads]{llncs}
\usepackage{graphicx}
\usepackage{verbatim}
\usepackage{booktabs}
\usepackage{multirow}
\usepackage{rotating}
\usepackage{subcaption}

\usepackage{textpos}
\usepackage{framed}

\usepackage{url}
\begin{document}

\title{How to Drill Into Silos: Creating a Free-to-Use Dataset of Data Subject Access Packages}

\titlerunning{How to Drill Into Silos}
\author{
	\textit{Research Paper} \\
Nicola Leschke\inst{1}\orcidID{0000-0003-0657-602X} \and
Daniela Pöhn\inst{2}\orcidID{0000-0002-6373-3637} \and Frank Pallas\inst{1}\orcidID{0000-0002-5543-0265}}
\authorrunning{N. Leschke et al.}
\institute{Paris Lodron Universität Salzburg, Salzburg, Austria \\
\email{firstname.lastname@plus.ac.at} \and
Universität der Bundeswehr München, München, Germany \\
\email{daniela.poehn@unibw.de}
}
\maketitle              %

\begin{textblock*}{1.5\textwidth}(-3cm, -9cm) %
\begin{center}
\begin{framed}
    \textit{This preprint (2024-07-04) has not undergone peer review or any post-submission improvements or corrections. The Version of Record of this contribution will be published in the\\ Proceedings of the \textbf{12\textsuperscript{th} Annual Privacy Forum (APF 2024)} \\ and will be available online at \\ \url{https://link.springer.com/conference/apf}.}
\end{framed}
\end{center}
\end{textblock*}

\begin{abstract}
The European Union's General Data Protection Regulation (GDPR) strengthened several rights for individuals (data subjects). One of these is the data subjects' right to access their personal data being collected by services (data controllers), complemented with a new right to data portability. Based on these, data controllers are obliged to provide respective data and allow data subjects to use them at their own discretion. 

However, the subjects' possibilities for actually using and harnessing said data are severely limited so far. Among other reasons, this can be attributed to a lack of research dedicated to the actual use of controller-provided subject access request packages (SARPs). To open up and facilitate such research, we outline a general, high-level method for generating, pre-processing, publishing, and finally using SARPs of different providers. Furthermore, we establish a realistic dataset comprising two users' SARPs from five services. This dataset is publicly provided and shall, in the future, serve as a starting and reference point for researching and comparing novel approaches for the practically viable use of SARPs.

\end{abstract}

\section{Introduction}

The General Data Protection Regulation (GDPR)~\cite{gdpr} was implemented to streng\-then and harmonize data protection within the European Union (EU). Besides other principles \cite{the_piece_2024}, it particularly clarifies and details,
as per Art. 15, the data subjects', i.\,e., individuals', right to request information about how personal data concerning them are being processed and used. In addition, data subjects have the right to request a copy of their data from data controllers by performing data subject access requests (DSARs). 
The information has to be provided in an electronic and understandable format within a month and free of cost, with few exceptions. 
Similar to this electronic copy, data subjects have the right to receive a machine-readable copy of the data \emph{they provided} according to Article 20 GDPR (data portability). Even though both Articles include the right to get a copy of personal data and are therefore subsumed under the term ``ex-post transparency'' \cite{murmann_tools_2017}, the content and format of those copies differ from a legal point of view. Nevertheless, the interfaces found in practice do not necessarily make such a difference \cite{bufalieri_gdpr_2020}. Therefore, we define the electronic data received via a respective ex-post transparency request as \emph{subject access request package (SARP)}, which can be a copy of personal data according to Art. 15, a data portability archive according to Art. 20, or a general SARP according to alike regulations.

As one of the most comprehensive legislative changes introduced with the GDPR, the strengthened right to data access combined with the newly introduced right to data portability have been broadly studied in matters of their effects on users and services \cite{urban_study_2019,kroger_how_2020,veys_pursuing_2021}. Compared to related obligations regarding transparency \cite{gerl_lpl_2018,gruenewald_tilt_2021,grunewald_enabling_2023}, however, explicitly technical contributions concerning Art. 15 and 20 GDPR are rather rare. 

There might be different reasons for this, particularly including a lack of publicly available datasets of SARPs as the basis for conducting respective research.
In practice, such datasets can only be generated through access requests from actual accounts filled with realistic data. The datasets do, in turn, inevitably include personal data and using them for research purposes is only possible after properly de-identifying them. 
Still, a risk of re-identification remains~\cite{ramachandran_exploring_2012}, especially when SARPs corresponding to the same data subject but originating from different controllers are published.
Therefore, researchers either publish only selected information~\cite{boeschoten_port_2023}
or decide to keep the SARPs private~\cite{10.1145/3600160.3605064,sorum_dude_2021,urban_study_2019}.

To overcome these issues, the overall goal of this work is to provide a method to generate publicly available SARP datasets spanning both multiple data subjects and multiple controllers, thereby facilitating further research towards technical contributions to ex-post transparency. Beyond that, SARP datasets might also fuel research in other, non-privacy related areas, like social network analysis~\cite{zannettou_analyzing_2024}.
Hence, we discuss the method to generate SARPs based on research-only accounts and de-identification, create a corresponding dataset, and compare our approach with related work.
Our contribution is therefore two-fold: 
\begin{enumerate}
    \item We propose a method to receive SARPs and curate them into a dataset specifically tailored to answer research questions within different areas and 
    \item We provide an initial, minimal dataset spanning two data subjects and comprising 
    particularly relevant services, namely Apple, Amazon, Facebook, Google, and LinkedIn.
\end{enumerate}

These contributions unfold as follows: First, in section~\ref{sec:goals}, we present the goals of our work. 
Next, in section~\ref{sec:method}, we outline and discuss our method to create and fill the accounts, as well as to de-personalize the SARPs before the analysis. In section~\ref{sec:dataset-creation}, we subsequently describe the actual creation of the datasets. We discuss related work in Section~\ref{sec:related-work} before we conclude the paper.

\section{Goals}
\label{sec:goals}

DSARs and their resulting SARPs should allow data subjects to get a deeper understanding of the processing and, thereby, foster informed privacy decisions, which are a core concept in modern privacy legislation. In practice, however, the underlying rights are underused and not perceived as useful \cite{syrmoudis_unlocking_2024,veys_pursuing_2021}. To combat this lack of usefulness, there is an inherent need for research in the area of DSARs, ranging from conceptual research to the design and development of user-centric tools and applications \cite{10.1145/3600160.3605064,syrmoudis_unlocking_2024,gomez_personal_2024}. 
Moreover, researchers from other domains like social network analysis \cite{wei_what_2020,Boeschoten2021,zannettou_analyzing_2024}, or even (mental) health research \cite{hafen_personal_2019,verbeij_happiness_2024} are increasingly interested in using SARPs.
  
One current hindrance for respective research is the lack of publicly available, realistic, and controlled reference datasets spanning multiple controllers,  which could serve as a playground for development and as comparative baseline. Instead, most researchers resort to individual ad-hoc data acquisition
\cite{ausloos_shattering_2018,10.1145/3600160.3605064,syrmoudis_unlocking_2024}.
Generating such datasets is challenging due to, e.\,g., ethical and privacy-related considerations \cite{ohme_digital_2022,10.1080/19312458.2022.2109608,zannettou_analyzing_2024}, or the (un-) willingness of real data subjects to share their personal data for research purposes \cite{skatova_psychology_2019}. 

Having a public dataset that allows for the initial exploration of SARPs, in turn, could fuel research, especially in the privacy domain, but also beyond (see, e.\,g., studies on the subject of risk-based authentication~\cite{10.1145/3600160.3605024}). Once existing, such a dataset can be used for different research directions, ranging from subject-centered research (e.\,g., focused on analyzing behavior over time) and policy-focused questions (exploring, for instance, the impact of regulatory changes or data interoperability as a prerequisite for a useful right to data portability) to social science \cite{Boeschoten2021}, mental health \cite{verbeij_happiness_2024}, or geoinformatics \cite{hanny_clustering_2024}. Similarly, research focused on gaining insights into the operation and behavior of controllers (see Cambridge Analytica~\cite{8436400}) or on reverse engineering the inherent schema of a controller (e.\,g., to facilitate completeness checks or as a prerequisite for data interoperability) would also benefit from a properly shaped realistic and publicly available SARP dataset.

Against this background, we strive for the following goals to be achieved by the data generation method and the respective initial dataset proposed herein:

\begin{itemize}
    \item \textbf{G1: A public, free-to-use dataset of SARPs.} In order to facilitate future research in the area of DSARs, we identify the need for a publicly available dataset containing exemplary SARPs. 
    The dataset shall be free-to-use for anyone and any purpose. 
    Therefore, real personal data is unsuitable, as it would be subject to the GDPR and might only be processed for specific purposes.
     
    A so-shaped dataset will, on the one hand, lower the entry barriers for researchers by providing an initial overview and allowing for the exploration of different research questions referring to different aspects of DSARs. On the other hand, the envisioned dataset shall also serve as a reference to compare different SARP-related approaches and applications (e.\,g., personal information management systems, privacy dashboards, or data portability concepts like the Data Transfer Project\footnote{\url{https://github.com/google/data-transfer-project}, last accessed 04/10/2024}).
    
    \item \textbf{G2: Machine-readable data.} Automation is at the core of the envisioned processing of the SARPs and the respective research to be fostered.
    This calls for machine-readable (preferably structured or semi-structured) data.\footnote{Potential options for data extraction from unstructured SARP data -- e.\,g., based on advanced AI/ML approaches -- are considered out of scope herein and shall thus not be considered further.}
    \item \textbf{G3: Detailed data.} The SARP data to be provided shall be as detailed and complete as possible to unlock a broad variety of analyses and facilitate the application of a broad variety of research approaches. This shall benefit privacy-related analyses (e.\,g., to assess the completeness of the data) but also detailed examinations beyond the privacy domain.
    \item \textbf{G4: Realistic data.} The data themselves should be as realistic as possible. In particular, they should resemble all characteristics, peculiarities, and possible intricacies to be expected for real-world SARPs as provided by relevant controllers. Otherwise, they would miss crucial aspects, potentially leading to unusable tools and misleading research.
    \item \textbf{G5: Controlled data.} To allow for the sound assessment of the results from automated analyses, to lower the risk of re-identification after publication, and to limit the amount of multi-party data\footnote{Data concerning multiple data subjects, which might not be willing to share their data for research purposes.}, the data should be dedicatedly generated in a controlled fashion, with as little uncontrolled factors as possible.

    \item \textbf{G6: Two-dimensional analysis of the dataset.} 
    Specifically, we strive for a dataset that enables both \textbf{(6a) cross-controller} and \textbf{(6b) cross-subject} analysis. Unlocking the real value from the right of access (RtA) for data subjects calls for the possibility of integrating SARP data from different controllers and analyzing them conjointly (6a). At the same time, researchers (e.\,g., from social sciences) are interested in analyzing patterns within a specific controller but across subjects (6b). Both types of analysis and respective research shall thus be supported by the dataset.\footnote{Last but not least, this will also contribute to the development of tools and respective research in the field of data portability through, e.g., the reverse engineering of controller-specific data schemas and respective matching concepts.}
    \item \textbf{G7: A method to facilitate the repeated creation of such datasets.}
    Simply providing \emph{one fixed} dataset will hardly suffice for a wide variety of possibly intended SARP data analyses. Long-term studies (e.g., for measuring the effect of new regulations), as well as foreseeable changes on the side of data controllers, might, for instance, call for repeated data collection. In contrast, topic-specific studies may necessitate the inclusion of particular (sets of) controllers. To support such additional dataset creations and relieve respective researchers and developers from repeatedly having to identify crucial challenges and respective countermeasures, we also intend to explicate our dataset creation method for flexible reuse.
\end{itemize}

\section{General Method and Precursory Considerations}
\label{sec:method}

Recently, and closest to our work, Boeschoten et al.~\cite{boeschoten_port_2023} introduced a workflow and corresponding software to enable the automated collection of SARPs ``donated'' by real users. In order to protect the privacy of those users, Boeschoten et al.~\cite{laura_boeschoten_2021_4472606} cleaned the data before usage as the participants shared their full name, phone number, and email address. While this approach is useful in, for example, social science research to meet \textbf{G3} and \textbf{G4}, it does not properly address the remaining goals pursued herein.\footnote{In particular, the data is not controlled by the researchers (\textbf{G5}). Additionally, the re-identification risks increase when adding multiple sources, so \textbf{G6a} can better be fulfilled with non-personal accounts.} We, therefore, provide a similar method, albeit with the difference of not using real data of natural persons, but rather employing pseudonymous accounts specifically tailored for research.

In particular, our approach rests on the following precursory considerations:

\begin{itemize}
    \item \textbf{C1: Research-only Accounts.} 
    Within such research-only accounts, there are different levels of artificial creation and usage patterns: real data, pseudo\-nymized, or entirely artificial. 
Accounts with real data might give the most realistic SARPs. Having several accounts with actual data at one provider might have implications, such as data aggregation, inference with personal contacts, or a restriction of the number of accounts in terms of services (ToS). Pseudonymized accounts reduce the linkage to the actual data. Further identifiers could be included within the derived and observed data, so a complete pseudonymization might not be possible. Based on our experience, artificial intelligence (AI) pictures are promising, though services might try to combat these types of pictures. Profile generators provided inconsistent data when the profile data was non-US-based. If the account is deactivated, reactivating it might be more challenging. Even if the profile does not impersonate a natural person, fake accounts might be against the ToS. Therefore, depending on the controllers and the aim of the study, dedicated research-only accounts with real data of the researchers might be the only option. In those cases, the data has to be monitored closely and extensive data cleansing has to be done before publishing.    
In general, we recommend using pseudonymous accounts. To avoid detection, respect the terms of the controllers, and satisfy \textbf{G4}, we stay close to the truth.
    
    \item \textbf{C2: Data Provided.}
    The datasets received by GDPR Article 15 or Article 20 DSARs should contain all the data subjects \emph{provided} beforehand. In addition, requests according to Article 15 shall also contain \emph{derived and observed} data but do not have to (but sometimes nonetheless might) be provided in a machine-readable form (see \textbf{G2}). 
Automated post-processing or analysis, as ultimately envisioned herein, can, in consequence, only be guaranteed for subject-provided data obtained via Article 20.

This subject-provided data can be controlled (\textbf{G5}) during account creation and usage in terms of amount and type of data shared, and their completeness in the SARP can be assessed. However, to allow for meaningful (legal) analysis, the data derived and observed holds high value. 
Therefore, we tend to use Art. 15 requests, but choose machine-readable options where available.
    
    \item \textbf{C3: Datasets of Different Sizes.}
    Depending on the use case, the appropriate dataset size differs. For making generalizations, large datasets are required. Also, for machine learning applications, as used, for example, in \cite{hanny_clustering_2024}, big datasets are mandatory to avoid overfitting \cite{prusa_effect_2015}.
    In contrast, small datasets are well-suited for human exploration. 

    The method proposed hereafter should thus allow for the creation of small, human-understandable datasets but should also facilitate large-scale data generation. This can be achieved in three different dimensions: adding more controllers, adding more users, or repeating the access request periodically.
    
\end{itemize}

\begin{figure}[t]
\includegraphics[width=\textwidth]{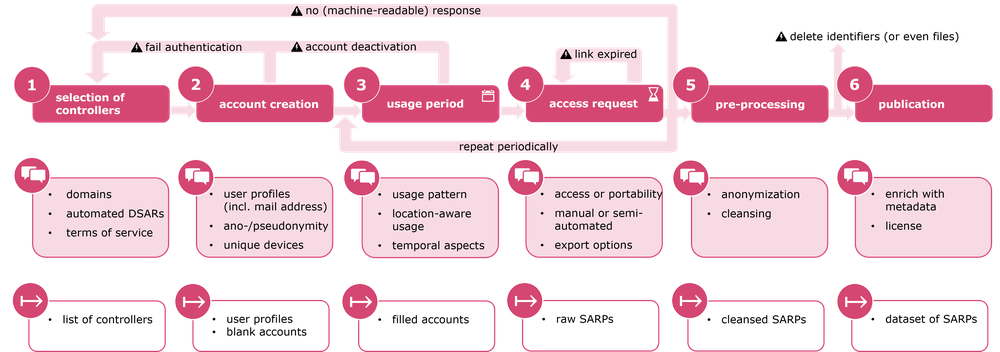}
\caption{Method for creating a dataset of SARPs.}
\label{fig:method}
\end{figure}

Building on these precursory considerations, we outline our method to generate a dataset containing a sample of SARPs from research-only accounts. The method -- visualized in Fig.~\ref{fig:method} -- contains six steps, each with specific necessary decisions, potential failures, and resulting outputs, as delineated below.

\subsection{Selection of Controllers}
\label{step1}
In general, the selection of controllers that should be included in the dataset depends on the research question but also has some implications on a meta-level: 
First, there are two important properties of the controllers to be examined, independently from the research question: a controller has to be subject to (one or many) DSAR obligations because of processing personal data and being subject to regulations like the GDPR, CCPA, or the like, and the DSAR processing should be automated. The latter is crucial in order to not raise unnecessary burdens to the controllers \cite{syrmoudis_unlocking_2024}.\footnote{The likelihood of a well-defined and automated access process is higher for big companies.} Moreover, it is important to validate the legitimacy of pseudonymous usage in terms of services and (depending on the study) possibly overriding national regulations.
The number of controllers, which we will refer to as $n$, and their distribution across domains should be chosen according to the specific research question(s) to be answered by the dataset.

In the second phase, a list of controller candidates is curated based on the research question(s) and the above-mentioned properties.
This list is narrowed down by an initial exploration of both the request process and the SARPs to be expected for getting a first impression of the suitability of the dataset. This can be done by analyzing related datasets or using the personal accounts of the researchers. The request process should be evaluated based on scientific reports or other sources (privacy policy, third-party instructions, exploration using personal account), assuming that there is a web-based request process and that the controller-sided process is automated as well \cite{10.1145/3600160.3605064}.
The resulting candidate list should be discussed in terms of suitability for answering the research question. Then, the fixed number of $n$ controllers is chosen from the candidate list. 

After finishing the step, there should be a list of $n$ suitable controllers to be included in the study. Possible failures in this phase occur when there is not enough information for the initial exploration of the request process and expected outcome. Additionally, this step might also be a good point to get feedback from an ethics commission.

\subsection{Creation of Accounts}
\label{step2}
Given the list of controllers, the next step is the creation of user accounts.
As outlined before, and to meet \textbf{G5}, our method is here specifically tailored for pseudonymous research-only accounts.
Therefore, within this step, the first action item is the creation of user profiles. 
This user profile has to include identification data, specifically a name, and, as the creation of an online account mostly relies on email addresses, an email alias, which should both be pseudonymous. In order to pass automated verification and depending on the domain, there might be the need to provide a credit card. This will have to be real data. Even more, some controllers require a unique phone number or even uniqueness of devices. Hence, there might be a need for a dedicated device and/or SIM card. Some controllers might also ask for physical addresses, which have to be within the territorial scope of meaningful access rights, e.\,g., the EU or California. Other profile data, such as date of birth, sex, education history, working history, or domain-specific information, can be entirely artificial. Overall, this makes the user profile as a whole pseudonymous and not entirely artificial.

As laid down above, one email address per user profile is required for identification purposes. Hence, the first account to be created is an email account. Using this email account, one account per controller and user profile is initiated.

As a result, $m$ user profiles and $n*m$ blank accounts are created.
If the user profiles fail the account verification of a specific controller, one should step back and choose another controller in Section~\ref{step1}. 

\subsection{Usage Period}
\label{step3}
Within a well-defined usage period, the blank accounts are filled with data by using the controllers' services as normally and realistically as possible.
Like creating accounts, data can be generated with real, pseudonymized, or artificial data. As discussed before, the variants have different implications. One exception is multi-party data, such as in private messages and chats. Here, the privacy of other people has to be taken into consideration. Options include chatting with other participating accounts or alternatively faking chats using, for example, AI text-generation tools.

Users may use services differently, ranging from the device (smartphone or notebook, different operating system, and browser) used for accessing to usage patterns (for example, before/after work and privacy-concerned) and frequency (low frequency vs. power user). Different types of usage may be better suited depending on the research questions.
The practicability for the researchers is another issue. For example, a smartphone may require a SIM card or can only be used with WiFi. 
In order to prevent too many ties to a natural person, usage is recommended only in public networks and not when traveling in private life. This usage pattern may result in the risk of forgetting to use the accounts, so implementing reminders is an important part of the study.
This is especially important when the research question requires a long usage period, for example when recurring DSARs are to be performed to allow for the analysis of SARP development.

At the end of the usage period, the accounts are filled with usage data. 
However, during usage, an artificial account might be detected and deactivated. In that case, real verification data is beneficial. If the account can not be reactivated, return to step 2.

\subsection{Data Subject Access Requests}
\label{step4}
To receive a SARP containing the usage data, the next step deals with DSARs.
 As mentioned above, the controllers within the study should allow for web-based DSARs. Still, the process of executing DSARs is often tedious and error-prone \cite{bowyer_human-gdpr_2022}. Leschke et al.~\cite{leschke_dara_2023} automate the user side of the request process, an approach that can prove beneficial for more extensive studies. After sending the request, researchers should regularly check the progress of the DSAR. Although the compilation of SARPs can take up to 30 days, it is often much faster, especially with automated processes.
 This unpredictable time frame is particularly challenging, as some controllers provide the SARPs only for a few days. If that timespan is passed, a new DSAR has to be triggered \cite{10.1145/3600160.3605064}. 
In order to meet \textbf{G3}, the DSAR should be based on Art. 15 GDPR (``data access''), as those might hold substantially more data than a portability request according to Art. 20. However, some web forms do not allow one to choose between different legal bases.
Often, users are allowed to choose between several request options, e.\,g., a timeframe, categories of data, or even the format of the SARP.

As a result, the researchers have gathered $n*m$ SARPs that contain account and usage data. 
However, there are still some controllers that are not responding to access requests or provide no machine-readable data, some even send a paper copy via mail. In this case, a DSAR based on Art. 20 should be used. Only in worst-case scenarios, one has to go back to step 1.

\subsection{Pre-Processing and Cleansing}
\label{sec:cleansing}
As hinted with \textbf{C1}, the data exports may contain personal identifiers and other possibly identifying data, such as IP addresses, geolocations, phone numbers, and names. 
Boeschoten et al.~\cite{Boeschoten2021} analyze the possibly identifying data (namely email, name, phone, URL, username, and faces) of Facebook, WhatsApp, Snapchat, and Twitter before presenting an approach for de-identification. However, the data and data structure may change, and the approach is limited to the websites presented. The amount of possibly identifying data depends on the preceding considerations and the level of pseudonymization. 
Pseudonymous data is vulnerable to re-identification attacks.
Hence, the data has to be pre-processed and cleansed before it can be analyzed.
By analyzing the data for possibly re-identifying information and using different methods to anonymize them, re-identification is made reasonably unlikely and the dataset can be assumed to be non-personal data \cite{finck_they_2020}.
A list of typical re-identifying information can be found in Table~\ref{tab:sensible-data}.
When other anonymization techniques fail, e.\,g., when having personal data within unstructured files like PDF and media files, the corresponding personal data or even file has to be deleted. This should be noted down and a list of deleted items should be provided as supplementary material.

\begin{table}[t]
\centering
\caption{Data with privacy- and security-related relevance.}
\label{tab:sensible-data}
\begin{tabular}{lll}
\toprule
\textbf{Type}            & \textbf{Privacy}    & \textbf{Security}      \\ \midrule
\multirow{6}{*}{Device}  & ID                  & ID                     \\
                         & Name                & Name                   \\
                         &                     & Type                   \\
                         &                     & Hardware               \\
                         &                     & Serial number          \\
                         &                     & OS version             \\ \midrule
\multirow{4}{*}{Network} & Type                & Type                   \\
                         & ISP                 & ISP                    \\
                         & IP address          & IP address             \\
                         & Location            & Location               \\ \midrule
\multirow{4}{*}{Website} & User ID             & User ID                \\
                         &                     & Sessions and cookies   \\
                         &                     & Authentication methods \\
                         &                     & User agent             \\ \midrule
\multirow{6}{*}{User}    & Name                & Account name           \\
                         & Email address       & Email address          \\
                         & Address             & Address                \\
                         & Birthday            & Birthday                     \\
                         & Phone number        & Phone number          \\ 
                         & Media               & Media \\
                         \midrule  
\multirow{3}{*}{Multiparty} & Usernames, User IDs    & Usernames, User IDs    \\
                         &   Timestamps              &    \\
                         &  Message content                   & Message content \\   
                         \bottomrule
\end{tabular}
\end{table}

However, further data can also call for proper cleansing. Concerning privacy, this particularly includes multi-party data -- such as messages and emails -- birthdays, addresses, internet service providers (ISPs), and user IDs. 
Further data included in the SARP may put the conducting researchers at risk in matters of security. For instance, information about device features may be used by adversaries to mimic the researchers' identity and try to attack them on that basis (see~\cite{10.1145/3372297.3417892}). Our list of to-be-cleansed data laid out in Table~\ref{tab:sensible-data} thus also includes data items without direct privacy-related relevance.
As a result, a set of cleansed SARPs that can be used for analysis and publication is provided.

\subsection{Publication}
It is important to provide supplementary information with the dataset to make it meaningful to the public. Particularly, the goals, methods, and pre-processing steps -- with a focus on anonymization techniques used and the list of deleted items -- are important additional information (see \textbf{G5}).

With the pseudonymous accounts, location-aware usage and dedicated pre-processing steps that are inherent within this method, the risks of re-identifica\-tion are minimized. However, as there are different attack vectors on datasets \cite{wong_minimality_2007}, and the risk of re-identification within social media data is high \cite{branson_evaluating_2020}, the risk cannot be ruled out. The original users (researchers) must be aware of the risks and benefits of publication and should, therefore, explicitly consent to its publication for all research purposes. 

The concrete publication modalities should be discussed. The options include providing the dataset for public download or making it available upon request. While the former removes all barriers to research, the latter lowers the risks of misuse.

\section{Initial Application: A Minimal SARP Dataset}
\label{sec:dataset-creation}

Following our method, we create a dataset to meet our goals defined in Section~\ref{sec:goals}. We decide on a micro-dataset to allow for manual exploration. Hence, the number of controllers and user profiles should be minimal but sufficient to reach goal \textbf{G6}. Addressing \textbf{G6a}, we set the number of controllers to five. A meaningful cross-controller analysis should cover both, controllers within a specific domain (that intuitively should have similar data) and controllers from different domains (to explore domain-independent data). 
To meet \textbf{G6b}, we include two user profiles.
In the following, we present the specific choices that were made to create such a dataset using the method presented in Section~\ref{sec:method}.

\subsection{Selection of Controllers}
We concentrate on the major online services of Google, Apple, Facebook, and Amazon (GAFA), as well as LinkedIn as a business social network. All of these are available for citizens within the EU and, thereby, subject to the GDPR. As major services, they are all viewed as gatekeepers according to the Digital Markets Act (DMA) and, therefore, of particular interest to researchers and the general public.

Previous work \cite{tolsdorf_case_2021,10.1145/3600160.3605064,syrmoudis_unlocking_2024} indicates that those controllers meet \textbf{G2} and offer web-based access requests \cite{leschke_dara_2023}, which lowers the burden on the controller (see Section~\ref{step1}).
With Facebook and LinkedIn, there are two services from the social network domain. 
Google serves as the e-mail provider, but is also an active player in several domains, of which we choose video streaming (via YouTube).
Within the same domain, we use Apple and Amazon as video streaming services. To demonstrate their inherent complexity, the former is additionally viewed as an app store provider, while the latter is used for online shopping.

\subsection{Creation of Accounts}
We created two pseudonymous user profiles, which we refer to as user $A$ and $B$, and respective research-only accounts for the selected websites.
While both profiles can overall be viewed as pseudonymous, especially verification data had to have some connection to real data, which we monitored closely in order to de-identify them during data cleansing (as described in Section~\ref{step4}).

Using the information of the user profiles, two Gmail accounts where opened first. Next, each Gmail address was used to generate an Apple ID. To preserve the uniqueness of devices and use Apple's app store, we used the iOS apps of the covered services. Each user profile was bound to a dedicated iPhone that was solely utilized for this study. After installation, the Facebook, Amazon, and LinkedIn apps were used to create accounts based on the Gmail address and the corresponding user profile.

Wherever possible, we stuck to the truth, mimicking realistic data. For example, Apple required billing information for one of the created accounts. There, we used the university's address. For the other account, another email address (university email address) and phone number were good enough to pass account verification. In order to get realistic data (\textbf{G4}), as provided by real data subjects, we allowed cookies and other tracking methods where they did not contradict the pseudonymous usage, e.\,g., we allowed location-based services only while using the apps.

\subsection{Usage Period}
In order to keep the accounts both realistic (\textbf{G4}) and controlled (\textbf{G5}), we agreed on the information we wanted to share, the kind of interactions to be performed, and the close and frequent communication. During the usage period, we took notes about the privacy-relevant decisions that were made.

For social networks, we agreed on the following actions. We followed accounts of huge organizations, but allowed only interaction with non-personal accounts, meaning either legal bodies or other research-only accounts. Public posts are liked and might be commented on. Chats are allowed between research-only accounts. Images might be uploaded if they convey no personal data, e.\,g., show objects without humans or are artificially generated. For video streaming, we watched free videos with ads and interacted with them (i.\,e., pausing, fast-forwarding or -backwarding, and stopping). For online shopping, we mainly used the search function and added items to the wish list.
Similarly, the app store search function was used, and free apps were installed.

Overall, we paid special caution regarding multiparty data, particularly messages and emails. These should not be made public unless all parties know that the content will be used for this study. Thus, we interacted only with the other research-only accounts. During pre-processing, all incoming messages had to be reviewed in detail, and, when in question, be deleted. For all other data categories, we tried to fill the accounts with as much data as possible to get realistic data.

With the purpose of pseudonymous usage, we limited the usage to mainly the universities' WiFis and other public and open networks. Hence, the usage was restricted to office hours, leading to infrequent users. Nonetheless, with creating a dataset for (privacy) research in mind, this limitation was acceptable. The usage period for users $A$ and $B$ was two and four months, respectively. The characteristics of the accounts are summarized in Table~\ref{tab:users}.

\begin{table}[t]
\centering
\caption{Characterization of the users.}
\label{tab:users}
\begin{tabular}{p{4cm}p{7cm}}
\toprule
\textbf{Characteristics} & \textbf{Users}              \\ \midrule
Type of profile & Research only                                 \\
Usage pattern          & Infrequent usage                          \\
Websites                 & Amazon, Apple, Facebook, Google, LinkedIn \\
Privacy attitude & Allow tracking\\
Actions                  & Search, like, save, hear/watch, chat      \\
Device                   & iPhone    \\ \bottomrule                              
\end{tabular}
\end{table}

\subsection{Data Subject Access Requests}
\label{sec:instanciation:request}
After a few weeks of usage, we requested the corresponding subject access request packages via web interfaces. When given the choice, we chose to request the data according to Art. 15 GDPR. Where possible, we chose a machine-readable format (e.\,g., selecting JSON instead of HTML) to reach \textbf{G2}. In general, we decided to include as much data as possible, setting the export options to all data (time and category-wise) to address \textbf{G3}.

For successful requests, the SARPs were available within six days. 
In one case (Apple), the initial 
request failed for one account completely.\footnote{We did not receive a response within 30 days}
After 30 days, a second request was made, choosing only selected data. 
For the second account, the response was incomplete. 
LinkedIn provided one basic export very quickly (within 20 minutes), and a 'complete' export within two days. 
In general, a few issues appeared during the process, such as unusual waiting time, no notification email though the SARP was downloadable, and expired download possibilities.

\subsection{Pre-Processing and Cleansing}
As stated in Section~\ref{sec:cleansing}, possibly identifying data within the SARPs needs to be obfuscated or removed in order to de-identify the dataset.  While the labels of such identifying information vary across controllers, we hereafter refer to the general concepts that need special caution. An extensive list of all identifying data, their paths, and the methods that are used to de-identify them is provided with the dataset. 
Per user, we de-identified the data in the same manner across providers, e.g., replacing the original mail address with \textit{firstname.lastname@gmail.com}, to preserve implicit cross-provider linkages. Across users, however, we employed different de-identifications to reflect the variety of available approaches.

First, we looked at the content of machine-readable files, which are of particular interest for future work.
For string-based identifiers like name, username, and mail, we chose two approaches: user $A$s data was obfuscated by using semantic descriptions (like \textit{Firstname Lastname}), 
while the corresponding data of user $B$ was replaced by the common substitute \textit{***}.
Usage dates are preserved, but the birthdate was set to a random date between 1960 and 2000.
As the addresses provided were the addresses of an institution, this data did not demand obfuscation.
However, observed location data, like the usage area, are generalized (e.g., from district to city).
To preserve the variety of IP addresses found in the SARPs, within the data of user $A$, we replaced every unique IP address with a random IPv4 loopback address.\footnote{This allows the address to be in range 127.0.0.0 - 127.255.255.255}
To make the data more realistic, we replaced all the IP addresses of user $B$ with the static IP 77.24.117.117 that occurred in the dataset.
For multiparty data, such as chats within the Facebook and LinkedIn SARPs and emails in Google's SARPs, we preserve only the messages between the research-only accounts.
Other data is deleted to protect the sanctity of mail. 

Device-specific security-related information is also obfuscated. Here, we do not want to preserve the variety and, hence, performed the same steps for both users:
All numbers within device identifiers, like version and operation system, are replaced with \textit{0}. 
The internet service provider (ISP) listed in the dataset was replaced by \textit{unknown}.

\begin{figure}[ht]
\centering
\begin{subfigure}{0.45\textwidth}
\centering
    \includegraphics[width=\textwidth]{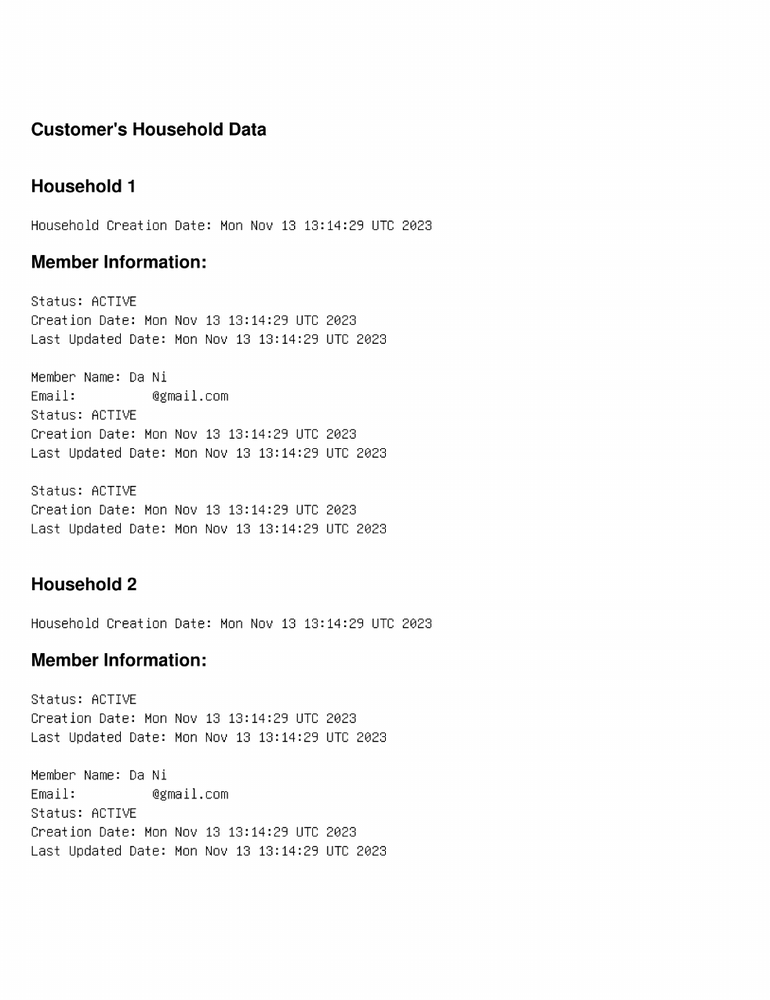}
    \caption{First page of the PDF.}
    \label{fig:amazon1}
\end{subfigure}
\hfill
\begin{subfigure}{0.45\textwidth}
\centering
    \includegraphics[width=\textwidth]{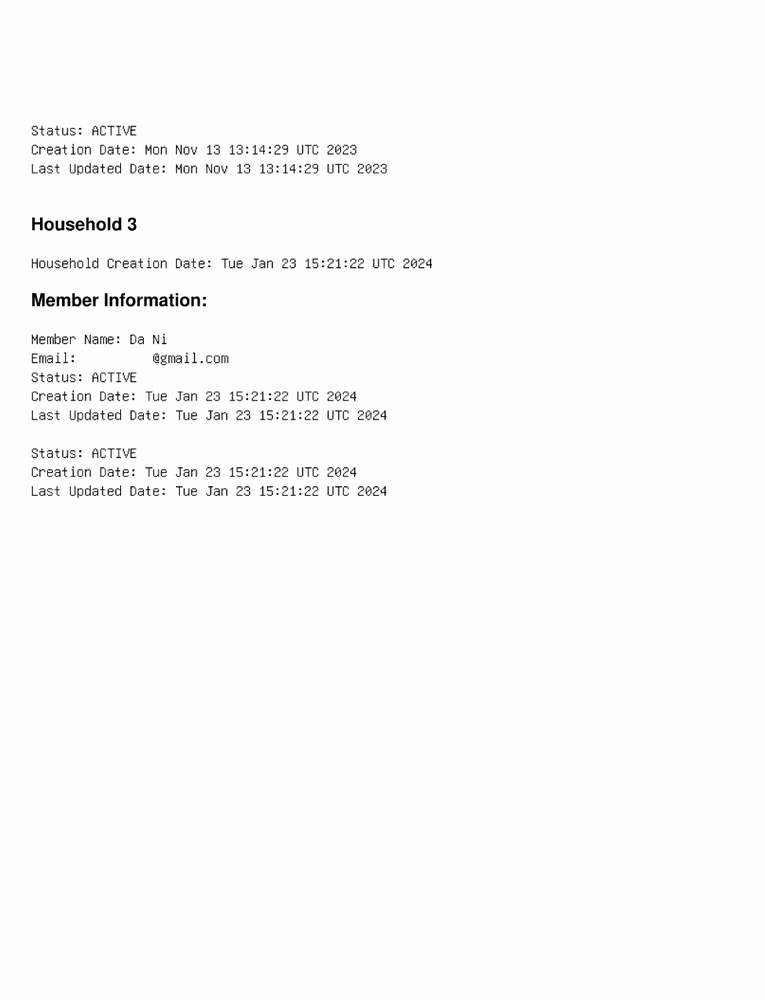}
    \caption{Second page of the PDF.}
    \label{fig:amazon2}
\end{subfigure}
\caption{Screenshot of two pages with information about the data subject's household provided by Amazon.}
\label{fig:amazon}
\end{figure}

Next, we looked at unstructured files. Media files, like pictures, could contain personal data. However, as the accounts were filled with the publication in mind, we never provided personal pictures and were also cautious with the metadata of the media files provided. We found personal information within PDF files provided by Amazon. Despite the fact that PDFs can be edited, the result may be reversed. Therefore, we decided to provide only screenshots of the edited content (see Fig.~\ref{fig:amazon}).

Finally, we examined the SARPs as a whole. Some folders and filenames included identifying information, e.\,g., the Facebook SARPs contained the username as the directory name. Here, we chose the same approach and replacement strategies as within machine-readable data. All de-identification steps are recorded and provided  with the dataset as supplementary material.

\subsection{Publication}
In order to allow meaningful analysis or development of user-centric applications, we enrich the dataset with metadata and provide a description of the method and its implementation within the dataset generation as laid down above. Additionally, an enumeration of files changed and screenshots of de-identified PDFs before deletion are provided. We make the dataset, including the cleansed SARPs and additional information, available for other researchers.\footnote{Access to the dataset can be requested through \url{https://doi.org/10.5281/zenodo.11634938}.}
The dataset is provided under a Creative Commons Attribution 4.0 license.

\section{Dataset Description and Preliminary Analysis}

In this section, we describe the dataset generated in Section~\ref{sec:dataset-creation} and perform a preliminary analysis for select examples to demonstrate its usefulness.

\subsection{Characterization of the Dataset}
A summary of the characterization of the SARPs for both users can be found in Table~\ref{tab:exports}. 
While usage time and export options denote important metadata to contextualize the files, the other columns describe the cleansed SARPs.
As stated in Section~\ref{sec:instanciation:request}, LinkedIn provided two separate exports, which we also describe separately. 
Apple and LinkedIn provide solely machine-readable data (particularly tabular data in CSV format). For Apple, this can be explained by a lack of unstructured data provided to the controller (no profile picture or similar), but on the other hand, it means that no metadata or explanations are provided. 
LinkedIn, in turn, provides download links for media files within one dedicated CSV file. Authenticated users must thus follow this link in a subsequent step to access this data.
For Amazon and Google, the ratio of machine-readable (semi-structured) files is around 70\% and for Facebook even less. This is due to media or domain-specific files (like calendar data as ICS) and contextualizing files like readmes.
Those contextualizing files are provided in the language chosen during usage (German in our case), while labels within semi-structured files are either in English or German (or abbreviations). Google also adapts folder and file names to the chosen language.

In general, the number of files may vary between subjects depending on the time and type of usage.
We observe that the file structure (number of files) is adapted accordingly.
Interestingly, the file type chosen within the export options also makes a difference (see Facebook).
The general file structure, precisely the varying depth across subjects, indicates that semantics/context is given not only within files but also in the folder structure. 

\begin{sidewaystable}[!htbp]
\centering
\caption{Characterization of the SARPs for subject $A$ and $B$.}
\label{tab:exports}
\begin{tabular}{lllllll}
\toprule
\textbf{Prov.}      & 
\begin{tabular}[c]
{@{}l@{}}\textbf{Usage time}\\ \textbf{(in months)}
\end{tabular} 
& \textbf{Export options}    & \textbf{File types}  & \textbf{\# Subfolders}   & \textbf{\# Files}         & \textbf{Export size}  \\ \midrule
Amazon   & 2 / 4    &  all categories                   &    \begin{tabular}[c]{@{}l@{}} 
CSV (32/49) \\ 
EML\footnote{Those files are deleted in the cleansing step, as they contain multiparty data. However, we included them in this list to demonstrate the different formats to be expected within SARPs. Same goes for MBOX and VCF files.} (2/5) \\  
JPEG (1/2) \\ 
JSON (3/3)\\ 
PDF (9/10) \\ 
TXT (4/4) 
\end{tabular}             &  41 / 49       &  51  / 73     &   1.2 MB / 1.4 MB               \\ 
\midrule
Apple    & 2 / 4    & \begin{tabular}[c]{@{}l@{}}
all data %
, \\ max. 1 GB / max. 4 GB \end{tabular} 
         & CSV (8/3)                 & 20 / 1            & 8 / 3            & 71.8 KB / 294.8 KB               \\ 
\midrule
Facebook & 2 / 4    & \begin{tabular}[c]{@{}l@{}}
all data %
, \\ JSON, \\ on my computer \end{tabular} & \begin{tabular}[c]{@{}l@{}}
JSON (39/0) \\ 
HTML (0/63) \\ 
TXT (29/28) \\ 
JPG (0/4) \\ 
PNG (1/15)\\ 
GIF (7/7) 
\end{tabular} & 45 / 76 & 76 / 117 & 12.3 MB / 13.5 MB               \\ 
\midrule
Google   & 2 / 4     & \begin{tabular}[c]{@{}l@{}}
all data, \\ frequency once, \\ ZIP, \\ 4 GB  \end{tabular} &  \begin{tabular}[c]{@{}l@{}}
HTML (8/11) \\
CSV (10/13), \\
JSON (27/28) \\
TXT (14/14) \\
PDF (1/1) \\
MBOX (1/1) \\
VCF (1/0) \\
ICS (1/0) \\
README (1/1) \\
JPG (0/2)
\end{tabular}
& 44 / 51  & 64 / 71  & 1.54\footnote{The MBOX file deleted has a size of 7.8 MB, which inflates the raw SARP.} MB / 1.2 MB  %
\\ 
\midrule
LinkedIn & 2 / 4   & all data         & CSV (18/21)   & \begin{tabular}[c]{@{}l@{}} 0 / 0 (part 1/2) \\ 0 / 0 (part 1/2) \end{tabular} & \begin{tabular}[c]{@{}l@{}} 13 / 18  \\ 19 / 21  \end{tabular} & \begin{tabular}[c]{@{}l@{}} 3.9 KB / 6.0 KB\\ 6.2 KB / 9.2 KB \end{tabular}  \\ 
\bottomrule
\end{tabular}
\end{sidewaystable}

\subsection{Preliminary Analysis}

To demonstrate its usefulness, we perform an exemplary manual analysis. Along this line, future research can use the provided dataset for exploration and as a baseline for evaluating user-centric applications for such SARPs. 
We select the topic of ads that were also targeted by Pöhn and Gruschka~\cite{apf-twitter} for Twitter (now X). Possible data comprises the targeting audience, shown ads, clicked ads, and the corresponding metadata (such as location that may hint at location-based ads). First, we analyze a single SARP by focusing on LinkedIn before comparing the analysis with the other online services of that specific user. This step demonstrates the dataset's suitability to reach \textbf{G6a}. Finally, to demonstrate \textbf{G6b}, we compare the results with the results of the second user.

The SARP of subject $B$ for LinkedIn consists of 21 CSV files, including 'Ad\_Tar\-geting', and 'Ads\_Clicked'. In 'Ad\_Targeting', several organizations are stated in the column of company followers. According to 'Ads\_Clicked', subject $B$ has clicked on 98 ads in four months, which seems too high for a user who tries not to click on ads. Unfortunately, only the ID of the ad is included in 'Ads\_Clicked'. Thereby, we do not know which ads were marked as clicked. The number of reactions, stated in 'Reactions', such as liking a post (which the user regularly did, following the profile filling strategy), is lower, with 30 in total.

Regarding the SARPs for Google and Apple, no ad-related data was found. This observation is interesting as ads were displayed while watching YouTube videos. In contrast, Amazon shows ``Interest based ads'' according to the file 'Advertising.OptOut.csv'. According to Amazon, nine ads were streamed within Amazon Music. However, no ads related to Prime Video could be found, although ads were shown according to subject $B$. Also, no data about clicked ads or the targeting group is included. This contrasts with Facebook, where the two interacted ads are shown, and the target group consists of sales and mead (sic!). However, Facebook does not include data about ads that they have shown.

When comparing the ad-related data of subject $B$ with those of subject $A$, we also noticed several ads for the latter, again with the amount of clicked (28) exceeding the one of reactions (11). Like for $B$, the file 'Ad\_Targeting' seems misaligned. For example, company industries have the entry of '2025'. In contrast to subject $B$, the member age and selected further data are not given. The most significant contrast between $A$ and $B$ can be found with Amazon's SARPs. While Amazon does not provide more detail on ads for subject $B$, it does so for subject $A$ with advertiser audiences (70 in total), advertiser clicks (3), and Amazon audiences (9 categories).

These (due to space constraints) rudimentary findings should already illustrate the kind of research and insights possible with the kind of datasets proposed and provided herein. Further, more detailed insights would go beyond the scope of the paper at hand and are planned for a subsequent publication.

\section{Related Work}
\label{sec:related-work}
Following the introduction of the GDPR, the CCPA, and similar legislation, DSARs gained significant research attention throughout recent years. Large portions of respective research so far regard real-world DSAR processes, ranging from their empirical examination and analysis \cite{borem_data_2024,dewitte_chronicling_2024,syrmoudis_unlocking_2024,10.1145/3600160.3605064,apf-twitter,tolsdorf_case_2021,urban_study_2019} to the identification of generalized user journeys ~\cite{pins_finding_2022}. Common findings include a limited understandability of provided data \cite{bowyer_human-gdpr_2022,veys_pursuing_2021} and difficulty in actually executing DSARs in the first place \cite{pins_finding_2022,petelka2022generating}. Proposals for addressing these include data dashboards \cite{raschke_designing_2018,bier_privacyinsight_2016,herder_privacy_2020,tolsdorf_case_2021} and personal data information management systems (PIMS) \cite{barreau_context_1995,whittaker_mood_2020,wilhelm_vision_2023}. With a more pragmatic focus on the actual givens, easing the DSAR execution through established means of web automation has also been proposed~\cite{leschke_dara_2023}. 
None of these endeavors, however, focuses on the content of the provided SARPs or the creation and provision of respective datasets.

SARPs themselves, in turn, have been analyzed from a single controller's point of view, e.\,g., Twitter (now X) \cite{apf-twitter} or Instagram \cite{Peters2023}, and from a comparative perspective \cite{urban_study_2019,kroger_how_2020,10.1145/3600160.3605064}.
SARP-focused user studies, in turn, repeatedly conclude that the provided raw data are hardly useful for data subjects~\cite{veys_pursuing_2021,borem_data_2024}. Instead, SARP data needs to be automatically pre-processed and analyzed –– also across different providers –– in order to be actually useful.
So far, however, respective measures for automated pre-processing, -analysis, and exploration are lacking. Insofar, our contributions provided herein complement said work and shall foster the kind of research that helps the above-mentioned research strands to move forward.

As for the use of SARP data in research, Habu and Henderson~\cite{habu_data_2023} describe how data subject rights can be used for user-fueled research.
For example, social media researchers increasingly rely on SARPs.
Boeschoten et al.~\cite{boeschoten_instagram_2021,laura_boeschoten_2021_4472606,10.5117/CCR2022.2.002.BOES,Boeschoten2021} propose the usage of donated (and de-identified) SARPs for (social science) research. 
Similarly, Razi et al.~\cite{10.1145/3491101.3503569,10.1145/3579522,10.1145/3579608} present a case study on collecting Instagram SARPs for adolescent online risk detection and Zannettou et al.~\cite{zannettou_analyzing_2024} analyze TikTok SARPs to observe the effect of TikTok's recommendations on user engagement.
Even though a different line of research than ours, such studies will clearly benefit from our data collection method and the respective clarification of necessary considerations in various ways.
At the same time, our work clearly builds upon generalized higher-level insights from such research, particularly with regard to ethical and procedural aspects ~\cite{10.1080/1369118X.2019.1627386,10.1080/19312458.2022.2109608}.

\section{Discussion}

Based on our method introduced herein, we generated a first minimal dataset suitable for researching and developing novel approaches in the context of data access research and for benchmarking different approaches against each other. Even though tailored to meet the goals outlined in Section~\ref{sec:goals}, our method has some limitations.

First, our approach consciously favors controlled data and cross-study value over data realness and volume, contrasting pre-existing approaches based on data donations from real users. More representative and extensive data might be favored depending on the intended use case. However, the pre-processing and cleansing steps then require more work. Still, data donations from real users will provide more voluminous and realistic data, albeit at the cost of controlledness and potential privacy risks laid out above.

Second, depending on the language settings of the device and the accounts, the SARP may contain data in different languages and sizes. Based on the languages and locations, the data may be different. Exploring respective effects and implications is, though, up for future work. In addition, the amount of data may vary between different DSARs. In our case, for instance, $B$ had initially (four months difference) received data from Apple that contained more folders and files, including account information. In addition, the name of the content-holding folder changed between the two respective SARPs. This shows that a long-term study might be interesting.

Lastly, the terms of usage may prohibit users from employing pseudonymized accounts for research. Especially for social media applications, pseudonymous accounts might qualify as fake profiles and may therefore be seen as deceptive usage \cite{guo_online_2021}. 
Respective rights possibly overruling such terms may vary between countries and decisions~\cite{10.1093/ijlit/eau002}. 
In this regard, the German Psychological Society, the Federation of German Psychologists, and the American Psychological Association declare that no studies based on deception are to be carried out unless deception techniques are justified, for example, by a significant gain in scientific knowledge, and no alternative procedure without deception is available~\cite{apa,psych}. In our case, the avoidance of any interactions with individual non-study users largely eliminates the risk of deceit. Thereby, this procedure can be seen as justifiable as new knowledge is gained.\footnote{Given the explicit focus on research related to GDPR-based DSAR, resorting to Art. 40 DSA is also not a reasonable option.}
Nonetheless, researchers may risk the deactivation of the research accounts.

\section{Conclusion}

Recent privacy legislation, such as the GDPR or the CCPA, has stimulated research in a broad variety of subjects, particularly privacy statements, data sharing, and consent practices.
Compared to these, technical contributions concerning ex-post transparency requests are rarely found. However, the process of the DSARs and the content of the SARPs can provide interesting insights into the behavior and compliance of data controllers and strenghten the data subjet's sovereignity. Therefore, we identified the need for publicly available datasets of SARPs that can fuel future research in this area.

One recurring issue with SARP-related research is that these packages include personal identifiers in various forms and at various locations. Hence, the data first needs to be de-identified before analyzing and publishing them in line with best practices for data-driven research. To facilitate such research, we proposed a method to create free-to-use datasets of SARPs based on research goals. We then applied our method to the creation of our exemplary dataset. After characterizing our dataset, we exemplarily analyzed the topic of ads in our dataset. 
Finally, we contrasted and discussed our approach with related work. 

In future work, we plan to extend our dataset by including SARPs from additional providers and countries of origin and by adding more users. We also plan to study SARPs in the long term to recognize changes.
The analysis should be extended, for example, 
by performing analyses for different topics or by exploring different approaches for automation. 
Further important aspects for future work are applications for SARPs that can be developed using the dataset provided herein.
For example, privacy dashboard development could be enhanced by taking our dataset as a use case.
More research in personal data interoperability is required to allow integrated dashboards combining SARPs of multiple controllers.
Our minimal dataset can be used to explore different data integration approaches on such personal data, which will also benefit research on personal data portability. 

Overall, we thus envision our method and minimal SARP dataset to foster research in the domain of data subject access requests in a multitude of ways.

\subsubsection{Acknowledgements.}
\begin{small}
This publication has been supported by the EXDIGIT (Excellence in Digital Sciences and Interdisciplinary Technologies) project, funded by Land Salzburg under grant number 20204-WISS/263/6-6022.
\end{small}

\bibliographystyle{splncs04}
\bibliography{references}
\end{document}